\documentclass[12pt]{article}
\usepackage{amssymb,epsf}
\def\lsim{\mathrel{\rlap {\raise.5ex\hbox{$ < $}}
{\lower.5ex\hbox{$\sim$}}}}
\def\gsim{\mathrel{\rlap {\raise.5ex\hbox{$ > $}}
{\lower.5ex\hbox{$\sim$}}}}
\def\sqr#1#2{{\vcenter{\vbox{\hrule height.#2pt
        \hbox{\vrule width.#2pt height#1pt \kern#1pt
           \vrule width.#2pt}
        \hrule height.#2pt}}}}

\def\lsim{{\displaystyle
{{\raise-8pt\hbox{$ <$}}
\atop{\raise5pt\hbox{$\sim$}}}}}
\def\gsim{{\displaystyle
{{\raise-8pt\hbox{$ >$}}
\atop{\raise5pt\hbox{$\sim$}}}}}
\def\slsim{{\displaystyle
{{\raise-8pt\hbox{$\scriptstyle <$}}
\atop{\raise5pt\hbox{$\scriptstyle \sim$}}}}}
\def\sgsim{{\displaystyle
{{\raise-8pt\hbox{$\scriptstyle  >$}}
\atop{\raise5pt\hbox{$\scriptstyle \sim$}}}}}
\newskip\humongous \humongous=0pt plus 1000pt minus 1000pt

\newcommand{\sumpf}[0]{\sum_{(H^{\rm f},G^{\rm f})}\! \! \! \!
{\raise
4pt
\hbox{$'$}}\,}
\newcommand{\sump}[0]{\sum_{(H,G)}\! \! {\raise 4pt \hbox{$'$}}\,}

\def\bs{\begin{subequations}}
\def\es{\end{subequations}}
\catcode`\@=11
\newcount\hour
\newcount\minute
\newtoks\amorpm
\hour=\time\divide\hour by60
\minute=\time{\multiply\hour by60 \global\advance\minute by-\hour}
\edef\standardtime{{\ifnum\hour<12 \global\amorpm={am}%
        \else\global\amorpm={pm}\advance\hour by-12 \fi
        \ifnum\hour=0 \hour=12 \fi
        \number\hour:\ifnum\minute<10 0\fi\number\minute\the\amorpm}}
\edef\militarytime{\number\hour:\ifnum\minute<10 0\fi\number\minute}
\def\draftlabel#1{{\@bsphack\if@filesw {\let\thepage\relax
   \xdef\@gtempa{\write\@auxout{\string
      \newlabel{#1}{{\@currentlabel}{\thepage}}}}}\@gtempa
   \if@nobreak \ifvmode\nobreak\fi\fi\fi\@esphack}
        \gdef\@eqnlabel{#1}}
\def\@eqnlabel{}
\def\@vacuum{}
\def\draftmarginnote#1{\marginpar{\raggedright\scriptsize\tt#1}}
\def\draft{\oddsidemargin -.2truein
        \def\@oddfoot{\sl preliminary draft \hfil
        \rm\thepage\hfil\sl\today\quad\militarytime}
        \let\@evenfoot\@oddfoot \overfullrule 3pt
        \let\label=\draftlabel
        \let\marginnote=\draftmarginnote
   \def\@eqnnum{(\theequation)\rlap{\kern\marginparsep\tt\@eqnlabel}%
\global\let\@eqnlabel\@vacuum}  }
\def\subequations{\refstepcounter{equation}%
  \edef\@savedequation{\the\c@equation}%
  \@stequation=\expandafter{\theequation}
  \edef\@savedtheequation{\the\@stequation}
  \edef\oldtheequation{\theequation}%
  \setcounter{equation}{0}%
  \def\theequation{\oldtheequation\alph{equation}}}

\def\endsubequations{\setcounter{equation}{\@savedequation}%
  \@stequation=\expandafter{\@savedtheequation}%
  \edef\theequation{\the\@stequation}\global\@ignoretrue
  \vspace*{-12pt} \\}
\def\bs{\begin{subequations}}
\def\es{\end{subequations}}
\relax
\def\Im{\,{\rm Im}\, }

\def\thefootnote{\fnsymbol{footnote}}
\def\be{\begin{equation}}
\def\ee{\end{equation}}
\def\ba{\begin{eqnarray}}
\def\ea{\end{eqnarray}}

\def\ee{\end{equation}}
\def\bea{\begin{eqnarray}}
\def\eea{\end{eqnarray}}

\def\np#1#2#3{Nucl. Phys. {\bf{B#1}} (#2) #3}
\def\pl#1#2#3{Phys. Lett. {\bf{B#1}} (#2) #3}

\def\pr#1#2#3{Phys. Rev. {\bf{D#1}} (#2) #3}

\newcommand{\uarrw}[0]{\mathrel{
{\raise.5ex\vbox{\hrule width 1cm}\hskip-6pt\rightarrow}}}
\def\thebibliography#1{%
\vskip 0.5cm \centerline{\bf References}
\list{%
[\arabic{enumi}]}{\settowidth\labelwidth{[#1]}
\leftmargin\labelwidth
\advance\leftmargin\labelsep
\usecounter{enumi}}
\def\newblock{\hskip .11em plus .33em minus .07em}
\sloppy\clubpenalty4000\widowpenalty4000
\sfcode`\.=1000\relax}

\renewcommand{\theequation}{\arabic{section}.\arabic{equation}}

\renewcommand{\section}{\setcounter{equation}{0}\@startsection%
{section}{1}{0mm}{-\baselineskip}{0.5\baselineskip}%
{\normalfont\normalsize\bfseries}}

\renewcommand{\subsection}{\@startsection%
{subsection}{2}{0mm}{-\baselineskip}{0.5\baselineskip}%
{\normalfont\normalsize\slshape}}
\begin{document}
\renewcommand{\theequation}{\arabic{section}.\arabic{equation}}
\begin{titlepage}
\begin{flushright}
HU-EP 00/40,\\
hep-th/0010086 
\end{flushright}
\begin{centering}
\vspace{1.0in}

{\bf \large Rank 48 gauge group in heterotic string}\ $^\dag$
\\
\vspace{1.7 cm}
{\bf { Andrea Gregori$^1$}} \\
\medskip
\vspace{.4in}
{\it  Humboldt-Universit\"at, Institut f\"ur Physik}\\
{\it D-10115 Berlin, Germany}\\

\vspace{2.5cm}
{\bf Abstract}\\
\vspace{.1in}
We discuss the existence of a non-perturbative gauge sector
that can raise the rank of the gauge group of the ${\cal N}_4=2$
heterotic string up to 48. These gauge bosons, that don't exist in six 
dimensions, co-exist with those originating from small instantons
shrinking to zero size.
\end{centering}
\vspace{4cm}
                                              
\hrule width 6.7cm
\noindent
$^\dag$\  Research supported by the ``Marie Curie'' fellowship
HPMF-CT-1999-00396.\\
\\
$^1$e-mail: agregori@physik.hu-berlin.de

\end{titlepage}
\newpage
\setcounter{footnote}{0}
\renewcommand{\thefootnote}{\arabic{footnote}}

\setcounter{section}{1}
\section*{\normalsize{\bf 1. Introduction}}

It is well known that, under certain conditions, 
the gauge group of the heterotic string, 
whose maximal rank is fixed, in the perturbative construction,
to be sixteen, can be non-perturbatively enhanced. 
This has been shown to happen,
for instance, in ${\cal N}_6=1$ compactifications of the $SO(32)$
string, when small instantons shrink to zero size \cite{w} 
\footnote{Extensions of the gauge group, that appear as non-perturbative from
the heterotic point of view, have also been observed in the M-theory
compactified on $S^1 \times T^4 \big/ Z_2$ in Ref. \cite{flo}.}.
This extension of the heterotic gauge group,
whose rank can be raised to 32, can be explicitly observed in
the type I dual construction \cite{gp}, where it appears perturbatively. 
Indeed, through the map to F-theory, configurations have been analyzed,
for which the gauge group can receive a bigger enhancement \cite{am}, 
but here we will consider only situations that we can study through
string-string duality, without passing through higher dimensional
theories.
Intrinsically, the existence of such non-perturbative states is
related to the nature of the space on which the  string
is compactified, so that it exists for any value of the coupling
constant. Indeed, from a geometrical point of view, in the 
case of the heterotic/type I dual pair mentioned above, 
the spaces on which the heterotic and the dual type I string are 
compactified are the same.
It is natural therefore to ask whether in certain cases 
a non-perturbative phenomenon such as that at work on the small instantons of 
the heterotic string, can provide a further enhancement of the 
type I gauge group,
that would eventually raised to a group of maximal rank $3 \times 16$.
By duality, such an extension should exist also on the heterotic side.
The aim of this note is to discuss this issue,
and to provide evidence for the existence
of such a further non-perturbative enhancement, in 
compactifications to four dimensions. As we will see,
the further compactification of the ${\cal N}_6=1$ theory
is essential for the appearance of new gauge bosons,
whose coupling does not depend on a volume but rather
on a complex structure modulus.

We will consider the problem both from the heterotic and the
type I point of view. Finally, we will discuss also the type IIA dual
point of view, commenting on a possible 
higher dimensional (M-theory) interpretation.
Our analysis ultimately provides a step toward the investigation
of type II/heterotic duality for ${\cal N}_4=2$ compactifications
of the type IIA string which are not realized on K3 fibrations.

\vskip 0.3cm
\setcounter{section}{2}
\setcounter{equation}{0}
\section*{\normalsize{\bf 2. The Heterotic and Type I string}}

We start by reviewing some facts about compactifications
of the heterotic string, obtained by toroidal compactifications
on $T^2$ of the ${\cal N}_6=1$ theory in six dimensions.
This theory has ${\cal N}_4=2$ supersymmetry.
The perturbative corrections to the couplings of the 
$F_{\mu \nu} F^{\mu \nu}$  
or $R^2$ terms of the effective action can be shown to have
a general form of the type:
\be
{1 \over g^2} \; \approx \; \Im S \, + \, \Delta (T) + \Delta (U) \, + \ldots
\, ,
\ee 
where $T$ and $U$ are the moduli associated to the K\"{a}hler class and
complex structure of the two-torus respectively \cite{dkl}--\cite{greek}.
The $\Im S$-term is the tree-level contribution, while
the functions $\Delta(T)$, $\Delta(U)$ appear at the one loop
\footnote{In the above formula, we give only the dominant behavior,
omitting any contribution of  Wilson lines and terms mixing
the contribution of these moduli.}.
When a type I dual orbifold exists, it can be shown \cite{ap} 
that the correction to the $F^2$ terms does not depend on $T$,
which on the other hand is mapped, under duality, 
into the modulus $S^{\prime}$. This field 
parameterizes the tree level effective coupling of the D5-branes sector
\cite{s}, namely the sector dual to the 
heterotic small instantons \cite{w}.    
The non-dependence of the effective coupling of the heterotic $F^2$
terms on this modulus is then a necessary requirement in order for the
duality to work, because on the type I side there is no
perturbative mixing of the two moduli in the effective gauge couplings.
On the other hand, the analysis of this duality tells us that
the heterotic field $T$, entering generically into the corrections
of various terms of the effective action, is not just a geometric
modulus, but indeed the coupling of a non-perturbative sector.
For generic heterotic ${\cal N}_4=2$ compactifications,
the dependence of the corrections on
this modulus can be therefore interpreted as the signal of
the running of states charged under both the perturbative and non-perturbative
sectors.

We may now ask how should we interpret the dependence of the corrections
on the other modulus, $U$. Can this be seen in some way as the
coupling of another, non-perturbative sector? And if yes, what is this
sector?
For sure, if $U$ has to be interpreted as the modulus parameterizing 
the gauge coupling of another sector, this latter cannot exist
in six dimensions: whatever could in fact be the coupling of this sector in 
six dimensions, toroidal compactification would then give him
a dependence on the volume of the torus, namely on the
field $T$, and not on $U$. It must therefore
necessarily be a sector that appears only \emph{after compactification}
from six to lower dimensions.  By duality, this must be true,
whenever it exists, also on the type I dual of the heterotic construction.
Indeed, the $F^2$ and $R^2$ corrections of the type I ${\cal N}_4=2$
effective action depend on this modulus \cite{ap,ms}, 
and we could ask the same question, namely what is the interpretation
we must give to the modulus $U$, also in the type I framework.

If we indicate respectively by $g_{(a)}$ and $g_{(b)}$ the couplings
of the small instantons and of this new, unknown, sector,
we would have:
\ba
{1 \over g^2_{(a)}} & \sim & \Im T \, ; \nonumber \\
{1 \over g^2_{(b)}} & \sim & \Im U \, .
\ea
When the heterotic torus $T^2$ is described by a product of two circles,
of radii $R_1$ and $R_2$, we would have:
\ba
{1 \over g^2_{(a)}} & \sim & R_1 R_2 \, ; \nonumber \\
{1 \over g^2_{(b)}} & \sim & R_1 \big/ R_2 \, .
\ea
The two couplings are therefore exchanged under T-duality along the second
circle. If we start from the $SO(32)$ heterotic string,
we can put on the second circle a Wilson line that breaks 
$SO(32)$ to $SO(16) \times SO(16)$, chosen in order to act 
as W in Ref. \cite{w}. T-duality along this circle
exchanges then the ``$SO(32)$'' with the 
``$E_8 \times E_8$'' theory, making clear that, while the modulus $T$
has to be interpreted as the coupling of the small instantons
of the $SO(32)$ theory, $U$ should be interpreted as the coupling
of the small instantons of the $E_8 \times E_8$ theory.
Our claim is that indeed, in four dimensions, both these
sectors are \emph{present at the same time}. What makes this possible 
is the fact that, unlike in field theory, in the toroidally
compactified heterotic string there is T-duality, with the
presence, at the same time, of momentum and winding states.
The ordinary small instantons correspond to non-perturbative objects
of the ``field theory'' part of the heterotic string, i.e. the one built
on  the Kaluza--Klein, momentum states, while the other non-perturbative 
states are pure stringy non-perturbative states. 
Indeed, as discussed in Ref. \cite{gh,sw},
zero-size $E_8 \times E_8$ instantons don't give rise to vector 
multiplets in six dimensions, but rather to tensor multiplets.
This fits with our interpretation of the field $U$: only after compactification
to lower dimensions the tensor multiplets give rise to vector multiplets.
If we decompactify the four dimensional theory to six dimensions
$(V_{(2)} \equiv R_1 R_2 \to \infty )$, we indeed observe the 
disappearance of this gauge sector from the effective action:
\ba
{1 \over g^2_{(a)}} & \to  & { \Im T \over V_{(2)}} \; =  \; 1 \, ; 
\nonumber \\
{1 \over g^2_{(b)}} & \to & { \Im U \over V_{(2)}} \; \approx \;
{ 1 \over R_2^2} \; \to \; 0 \, .
\label{4-6}
\ea
On the type I side, this is related to the fact that,
as explained in Refs. \cite{6danom},
the fixed points of the $T^4 \big/ Z_2 \sim K3$ compact space 
of the six dimensional ${\cal N}_6=1$ theory are associated to a gauge
bundle without vector structure, so that the 
non-perturbative states arising from small type I instantons
in six dimensions contain tensor multiplets instead of vector multiplets.
The appearance of new massless vector multiplets after
circle compactification from six dimensions was interpreted,
in Ref. \cite{gh}, from the heterotic point
of view, as due to the appearance of tensionless strings.
Here we want to stress that the appearance of these states
in the lower dimensional theory is not related to the actual
existence of a perturbative (subgroup of the) $E_8 \times E_8$ gauge group:
they are indeed present also in the $SO(32)$ theory.
The simultaneous presence of both the small instantons of
$SO(32)$ and $E_8 \times E_8$ is due to the T-duality of the heterotic 
string, and matches with the simultaneous presence of these states
on the type I dual theory, in which however only
one of these two sectors is non-perturbative.
Here however we are faced with a puzzle. If we go to six dimensions,
by decompactifying the two-torus of the heterotic string, and reinterpret
the vector multiplets of four dimensions in terms of vector and tensor
multiplets of six dimensions, we see that the number of these states
does not in general satisfies the constraint imposed by the 
vanishing of the six dimensional anomaly:
\be
N_H-N_V+29 N_T = 273 \, .
\label{6d}
\ee
The solution to this puzzle comes from the analysis of the type II
dual point of view.

\noindent

\vskip 0.3cm
\setcounter{section}{3}
\setcounter{equation}{0}
\section*{\normalsize{\bf 3. The Type II point of view}}

Since heterotic/type IIA duality in four dimensions
maps the heterotic dilaton field
into a perturbative, volume-form modulus of the type IIA string,
we expect that inspection of the latter can give a hint in understanding
what is happening.
On the type IIA string side, the
Horava--Witten orbifold of the M-theory is realized as an
ordinary K3 compactification, which admits an orbifold realization
as $T^4 \big/ Z_2$. This projection produces a twisted sector
dual to the $E_8 \times E_8$ gauge sector of the heterotic string.
A further $Z_2$ orbifold projection, besides the breaking
of supersymmetry to ${\cal N}_4=2$, produces also a new
twisted sector, corresponding to the $SO(32)$ sector. 
Indeed, the two orbifold projections are equivalent and 
interchangeable. Moreover, the $SO(32)$
and $E_8 \times E_8$ points are connected in the K3 moduli space,
and in type IIA only the the Cartan subgroup of the gauge group appears 
perturbatively. It has therefore no meaning to distinguish between 
$SO(32)$ and $E_8 \times E_8$: in the following we will simply refer
to ``rank 16'' factors of the whole gauge group. 
Only one of these factors appears perturbatively on the heterotic side.
The second one appears perturbatively only on the type I side.
There is however a third twisted sector, corresponding to the fixed points
of the product of the two $Z_2$ orbifold projections.
This gives rise to another rank 16 sector, which is non-perturbative
on both the heterotic and type I sides.
On the type IIA side, this $Z_2 \times Z_2$ orbifold has in total
48 fixed points, and it was repeatedly considered in the literature,
both in the framework of string \cite{gkr} or
of F-theory compactifications \cite{mvII,gm}.
It corresponds to the orbifold limit of the compactification
on a Calabi--Yau manifold with Hodge numbers $(h^{1,1},h^{2,1})=(51,3)$. 
Its perturbative spectrum contains 3+48 vector multiplets
and 4 hyper multiplets.
The perturbative corrections to the effective coupling of the $R^2$ term were 
computed in Ref. \cite{gkr}, and read \footnote{For simplicity,
here and in the following we will omit all the normalization coefficients
and the cut-off dependent term accounting for the infrared running. 
For more details about this, we refer the reader to Ref.~\cite{infra}.}:
\be
{1 \over g^2_{grav}} \; = \; - \log \Im T^1 | \eta ( T^1 ) |^4 -
\log \Im T^2 | \eta ( T^2 ) |^4 - \log \Im T^3 | \eta ( T^3 ) |^4 \, ,
\ee     
where $T^1$, $T^2$, $T^3$ are the moduli associated to the
K\"{a}hler classes of the three tori of the six dimensional compact space.
Since the compact manifold is not self-mirror, the above correction
is most probably modified by non-perturbative corrections.
However, the above expression already indicates us that,
under type II/heterotic duality, these moduli should be mapped into
the three moduli $S$, $T$, $U$ \footnote{We recall that, for large
$\Im X$, $-\log \Im X | \eta (X)|^4 \to 3 \Im X$,  
$-\log \Im X | \vartheta_4 (X)|^4 \to \log \Im X$.}.
From the type IIA analysis, it appears clearly that these moduli
parameterize the couplings of three \emph{equivalent} gauge sectors:
only one of them shares the ``bare'' coupling with the perturbative 
heterotic string, namely the one parameterized by the field $S$,
and an entirely perturbative heterotic dual exists only 
when the other two sectors
are not present. This happens if one of the two orbifold projections acts 
freely, in such a way that neither the corresponding twisted sector
nor the twisted sector corresponding to the product $Z_2 \times Z_2$
possesses fixed points. This freely acting orbifold corresponds to
the $CY^{11,11}$ manifold, which is a K3 fibration.
This construction, together with its heterotic dual, 
has been considered in Refs. \cite{mvII}--\cite{gkp}.
In that case, the corrections to the $R^2$ coupling read, 
on the type IIA side:
\be
{1 \over g^2_{grav}} \; = \; - \log \Im T^1 | \eta ( T^1 ) |^4 -
\log \Im T^2 | \vartheta_4 ( T^2 ) |^4 
- \log \Im T^3 | \vartheta_4 ( T^3 ) |^4 \, .
\ee     
The $\vartheta$ functions in the second and third term signal that
the corresponding sectors are massive.
We remark that, in general, it is not possible
to construct an heterotic model with a behavior of the coupling like:
\be
{1 \over g^2_{grav}} \; \sim \; \Im S -
\log \Im T | \eta_4 ( T ) |^4 - \log \Im U | \vartheta_4 ( U ) |^4 \, ,
\ee
namely, a model in which only the vectors corresponding to the
``SO(32)'' small instantons are present. Again, the substantial
``equivalence'' of the moduli $T$ and $U$ in the heterotic perturbative
construction is a consequence of T-duality of the heterotic string
\footnote{This duality can be broken by Wilson lines, but never in a way
to lift the mass of all the states of only one of the two sectors.}.
The example we provided was just intended to give the flavor of what is 
happening, without any aim to provide a concrete dual for a specific
situation. In general, the four dimensional ${\cal N}_4=2$ heterotic
string is dual to a type IIA string compactified on a K3 fibration ${\cal M}$.
One can easily realize that, under heterotic/type II 
duality, the heterotic torus $T^2$ is mapped inside the fiber on the type IIA 
side. In this space, there are always at least three two-cycles, that
must correspond to the three vector moduli present in any ${\cal N}_4=2$
heterotic compactification, namely $S$ (the base of the fibration)
and the two moduli of the torus, $T$ and $U$, in such a way that
$(\Im S \times \Im T \times \Im U) \approx {\rm Volume}({\cal M})$. 
The $T \leftrightarrow U$
symmetry of the perturbative heterotic string implies a corresponding
symmetry also for their images in ${\cal M}$, at least when the volume
of the base is large.
When on the heterotic side extra, small-instanton-like states appear, 
and the type IIA fibration degenerates, 
the singular points must appear in a symmetric way with respect
to the two cycles corresponding to the image of $T$ and $U$.
For what matters the six dimensional anomaly constraint,
we have to keep in mind that
the decompactification of the image of the heterotic two-torus 
is a singular limit  on the type II side, that leads in general
to what resembles a ``non-compact orbifold''.
The ``six dimensional'' theory obtained via this decompactification is not
therefore a true, smooth six dimensional theory, and there is no reason
to expect that the massless states satisfy the six dimensional anomaly 
constraint, Eq. (\ref{6d}) \footnote{For a further discussion, see 
Ref. \cite{poster}.}.

\noindent

\vskip 0.3cm
\setcounter{section}{4}
\setcounter{equation}{0}
\section*{\normalsize{\bf 4. Comments}}

It is natural to ask why these new states never appeared in the
various analysis of small instanton-like phenomena of the heterotic
string, and whether there can exist a description of them 
in terms of some effective theory of membranes or, in general,
solitonic solutions of some effective theory.
In order to answer to these questions, we must go back and see
\emph{how} the ``ordinary'' small instantons have been detected. 
These states are always described in terms either of 
D5-branes of the type I string, therefore using string-string duality,
or in terms of solitonic, or membranes, solutions
of ``field theory'' configurations that are supposed to contain
the heterotic string, such as those of M-theory.
However, in both these cases one always ``projects'' the 
underlying theory onto a subspace spanned by the only Kaluza--Klein states.
Namely, in order to give a description in terms of field theory
solutions, one looks for a geometric picture, that necessarily
discards a part of the states. In other words, one never has,
at the same time, one manifold and its T-dual. 
If the states of a certain kind of small instantons
(those of the $E_8 \times E_8$ or those of the $SO(32)$ string
as well) are described in terms of a field theory
built on geometric objects, like membranes, it is not possible
to describe, at the same time, also the T-dual states.

As we said in the previous section,
the new states correspond instead to ``true''
small instantons of the type I string, something very different
from what one usually calls the ``type I small instantons'', namely
the duals of the heterotic small instantons, the D5 branes.
Their existence can be inferred from the fact that, like the heterotic string,
also the type I string is compactified on a K3, with a bundle
with tensor structure.
However, the only way of explicitly observing all these states is through
the direct type II construction in four dimensions, as we did in the
previous section. There, T-duality of the heterotic string appears
as a symmetry in the moduli space, that maps, in the specific example
we considered, a twisted sector into another one. 
We argue that the place where these phenomena can be properly investigated
is the type IIA string construction in four dimensions \footnote{For a 
detailed analysis of certain specific cases, see Ref. \cite{stria}.}: 
these new states should appear at singularities of 
K3 fibrations, and their coupling would be derived from the
intersection of the corresponding cycles with those associated to the
moduli $T$ and $U$ \cite{al}. 
From this point of view, it is clear that, as we said, the type II fibration 
must possess a symmetry corresponding to the heterotic $T \leftrightarrow U$
symmetry, at least when the volume of the base is large.
This implies a symmetry in the fiber, so that 
the appearance of only cycles that ``intersect'' $T$ and not,
at the same time, cycles that ``intersect'' $U$ (or vice-versa), would
correspond to a breaking of the heterotic T-duality much stronger
than the one expected~\footnote{See for instance 
Ref. \cite{apl}.}, which is due 
to instanton phenomena, suppressed at the weak coupling (large $S$).
In this case, the breaking of T-duality would be ``explicit'',
for instance in the effective action corresponding to the massless
degrees of freedom, and persist even in the weak coupling limit.
We consider this phenomenon quite unlikely to happen, 
because it would contradict a perturbative symmetry of the
heterotic string even in the weak coupling regime.
To this regard, a comment is in order about certain
results of Ref. \cite{am}, for which it seems
that the number of tensor multiplets and the rank of the extra,
non-perturbative gauge group
one obtains at special points of the six dimensional heterotic theory,
give, after compactification to four dimensions, a result in contrast with
T-duality. More precisely, as discussed in section 6 of Ref. \cite{am}, 
the theory is only ``covariant'', but not
``invariant'' under T-duality (see discussion around Eq. (47) of that 
reference). The analysis is based on the duality
between the heterotic string compactified on a K3 and F-theory
compactified on a Calabi--Yau three-fold.
The existence of such a  duality was conjectured in Ref. \cite{mvI}, 
as a consequence of the decompactification
of both the four dimensional heterotic string compactified on
$T^2 \times K3$ and its type II dual compactified on a K3 fibration.
At the base of the argument, there was the conjecture of Ref. \cite{vafa}, 
according to which the type~II theory
is essentially F-theory compactified on a torus.
However, the chain of arguments leading from a string/string duality
in four dimensions to a string/F-theory duality in six dimensions 
presented in Ref. \cite{mvI} is in general not correct:
the passage from four to six dimensions is in fact very singular.
In order to get a well defined six dimensional limit, one has to assume 
in fact that the heterotic torus $T^2$ is mapped to a torus,
or a ${\bf P}^1$, on the type II side. 
However, in the examples of heterotic/type II dual pairs discussed
in Refs. \cite{gkp,gkp2}, where the map of the moduli of the two-torus
has been carefully taken into account, it happens that
the moduli $T$ and $U$, associated respectively
to the K\"{a}hler class and to the complex structure of the heterotic torus,
are mapped to two volume moduli of the fiber on the type II side, 
in such a way that $\Im T \times \Im U = V_{(4)}$, the volume of the fiber.
Something similar happens also more in general:
on the heterotic side both the K\"{a}hler class and the complex structure
of a two-torus belong to the vector multiplets. On the type II side, instead,
they belong one to the vector and the other to the hyper multiplets
(which one, it depends on whether we consider type IIA or type IIB).
It is therefore clear that the heterotic torus must map into something
in the fiber, but it cannot be a torus. In general, it is not a
${\bf P}^1$ either: both $T$ and $U$ must correspond to vector multiplets,
represented by two independent cycles in the fiber.
More precisely, they must map into two of the three cycles that are 
always present in any K3 fibration (see the lower bound of Eq. (30) 
of Ref. \cite{al}). 
In fact, in the case of duality with type IIA, together with the divisor dual
to the heterotic dilaton, they must span the ``volume'' of the 
Calabi--Yau manifold. The decompactification of the heterotic torus,
i.e. the limit $\Im T \to \infty$, is therefore something ill defined
on the type II side: if this decompactification is achieved
by passing through a five-dimensional theory, then also
$\Im U \to \infty$, or $\Im U \to 0$. In any case, even if one trades, 
via a bi-rational transformation, a ${\bf P}^1$, or a torus, in the fiber with 
one in the base, there remains a mismatch regarding what happens 
to the other cycle. The decompactification of the heterotic string
is a very singular operation on the type II side, that 
``destroys'' the initial K3 fibration of the type II string. 
Conversely, let's consider the F-theory compactified to six dimensions
on that K3 fibration, ${\cal M}$.
Now, either ${\cal M}$ is also fibered over a torus, or a ${\bf P}^1$,
that can be considered to correspond to the type II coupling, so that, in a
certain limit, this vacuum can be seen to correspond to
a (weakly coupled) type II vacuum, or is not. In the first case,
further compactification on a two-torus adds (at least) one
vector multiplet and one hypermultiplet. In the second case,
compactification on a torus must add at least one hypermultiplet,
in order to provide the coupling of the type~II string.
In any case, this compactification adds at least one hypermultiplet.
This means that, away from special points, it adds one vector and one
hypermultiplet.
On the heterotic side, instead, compactification of the six dimensional
theory on a two-torus adds, away from special points, two vector multiplets.
There is therefore in general a mismatch between the two theories.
The loss of track of the heterotic T-duality on the type II side
has its origin in the ill-definiteness of the decompactification.
These problems arise when we want to use an analysis based on six dimensional
anomaly in order to derive, via toroidal compactification,
the non-perturbative spectrum of the four dimensional theory. 
In this case, the (singular) Calabi--Yau space of the type II string 
cannot be used as a good compactification space:
the F-theory dual of the so derived six dimensional heterotic string
may well be compactified on a very different space,
that just ``projects'' on ${\cal M}$. Moreover, there is in general no
consistent field theory description accounting for all the massless
states at the same time. In the cases in which
one considers only the perturbative moduli space of the heterotic theory,
and translates its properties into geometrical properties of
smooth manifolds, implementing then the analysis with a set of rules based
essentially on field theory, this duality \cite{mvI} works,
and can be transferred from six to four dimensions without
problems: in six dimensions, the perturbative theory is in fact highly 
constrained, and the four dimensional theory is simply a dimensional reduction
of that.

\noindent

\vskip 0.3cm
\setcounter{section}{5}
\setcounter{equation}{0}
\section*{\normalsize{\bf 5. Conclusions}}

In this note we discussed the existence, in heterotic
and type I $T^2 \times T^4 \big/ Z_2$ compactifications to four dimensions,
of a non-perturbative gauge sector, that extends the whole gauge group to
a maximal rank 48. This sector originates from wrapped
tensor multiplets associated, on the type I side, to the $Z_2$ fixed points. 
On the heterotic side, these vector multiplets are generically
present together with those appearing when small
instantons shrink to zero size. The simultaneous presence of both
these non-perturbative sectors in the heterotic string
is essentially due to T-duality, that exchanges momentum states
with winding states. The coupling of these extra gauge bosons,
that don't exist in six dimensions, is parameterized by the field $U$,
associated to the complex structure of $T^2$.
The analysis we carried out is only qualitative, and we
leave for future investigation interesting issues like what really are
the allowed gauge groups and what is the massless spectrum,
including hypermultiplets. From the investigation of the
type IIA string we learn that, from the M-theory point of view, 
the three gauge sectors, namely the one corresponding to the heterotic 
perturbative gauge group and the two non-perturbative ones, should
appear essentially on the same footing.
By this we mean that physics cannot ``prefer'' one sector
with respect to the other. Only
the identification of the heterotic dual chooses a
preferred direction.
This fact may have interesting consequences for the string phenomenology:
there is in fact no a-priori reason to prefer one of the three gauge sectors
to be the one that should contain the Standard 
Model~\footnote{For a discussion of this issue, see Refs. \cite{bo,gshort}.},
and the relation between gauge couplings and
string parameters may be rather different
from what expected from a perturbative heterotic string analysis.
In this case, we have in fact three gauge sectors,
in which the coupling is parameterized respectively
by the fields $S$, the axion--dilaton field of the heterotic string,
$T$, the modulus associated to the volume of a two-torus,
and $U$, associated to its complex structure.
In particular, in the third sector the coupling does dependent 
neither on the string scale, nor on the size
of internal dimensions, but only on a ``shape'' modulus,
both for the heterotic and the type I string.


\end{document}